\documentclass[aps,superscriptaddress,amsmath,amssymb,twocolumn,amsfonts]{revtex4-1}
\usepackage{amsmath}
\usepackage{graphicx}
\graphicspath{{Pictures/}}
\usepackage{hyperref}
\hypersetup{pdfnewwindow=true, colorlinks=true, linkcolor=magenta, anchorcolor=blue, citecolor=red, filecolor=blue, menucolor=blue, urlcolor=blue}
\usepackage{cleveref}
\usepackage{color}
\usepackage{float}
\usepackage{soul}
\usepackage{ulem}
\usepackage{braket}

\newcommand{\vg}{\ensuremath{\mathsf{g}}}

\begin{document}

\title{Mesoscopic critical fluctuations}

\author{Saikat Banerjee}
\affiliation{Theoretical Division, T-4, Los Alamos National Laboratory, Los Alamos, New Mexico 87545, USA}
\author{Nikolai A. Sinitsyn}
\affiliation{Theoretical Division, T-4, Los Alamos National Laboratory, Los Alamos, New Mexico 87545, USA}

\date{\today}

\begin{abstract}
We investigate magnetic fluctuations of a mesoscopic critical region at the interface induced by smooth time-independent spacial changes of a control parameter across its critical value. Near the spatial critical point, the order parameter fluctuations are mesoscopic, i.e., much larger than the lattice constant but decaying away from the critical region.   
We propose a minimal model to describe this behavior and show that it leads to the integrable Painlev\'e-II equation for the local order parameter. We argue that known mathematical properties of this equation produce insight into the nonlinear susceptibilities of this region. 
\end{abstract}

\maketitle

\section{Introduction \label{sec:sec_I}}

\begin{figure}[t!]
\centering \includegraphics[width=\columnwidth]{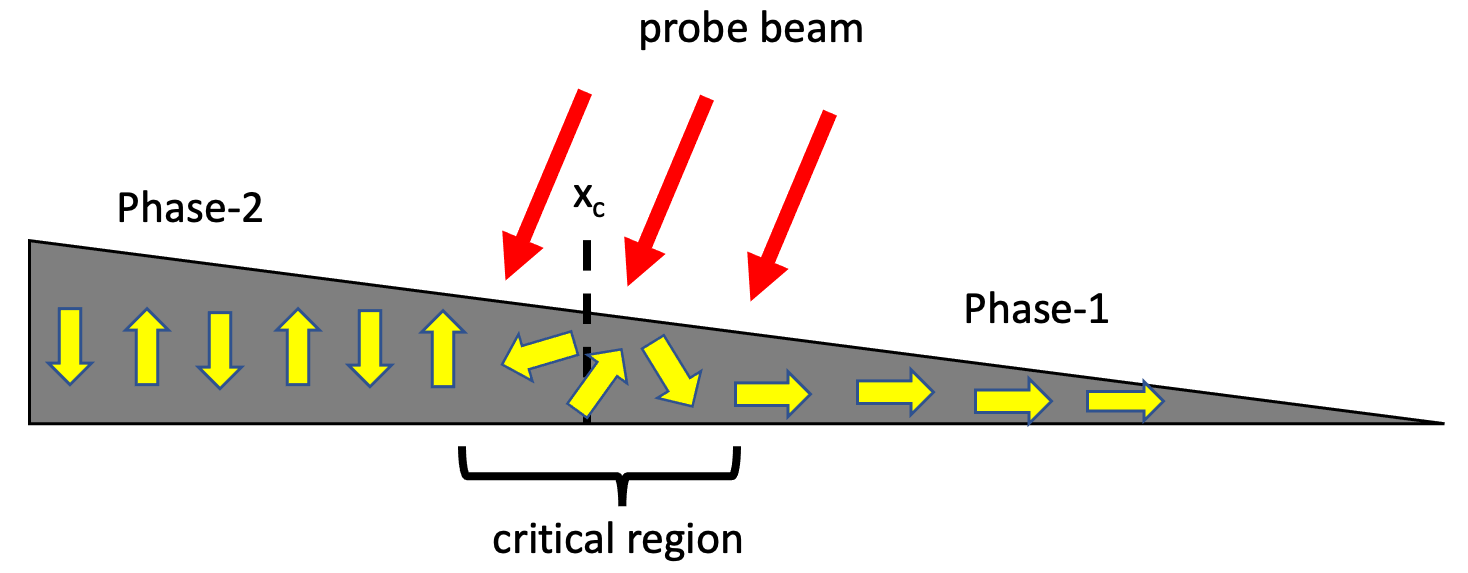}
\caption{A phase transition induced by a gradient of the width of a magnetic film. The critical point at $x=x_c$ separates two different magnetic phases. A mesoscopic region near this point is formed, which shows enhanced magnetic fluctuations, which are probed by an optical beam with a comparable width.}\label{pic-pt}
\end{figure}

This article is inspired by experimental studies of local magnetization fluctuations in a thin magnetic film with a {\it spatially} smoothly varying thickness \cite{PhysRevB.90.184404,Young2009,PhysRevB.74.064403,PhysRevB.71.104431,Shin2007,PhysRevX.8.031078}. Originally, this spatial variation was introduced in \cite{PhysRevB.90.184404} to guarantee the presence of a critical region. Thus, on the thin side of the film, the interactions favored the phase with an in-plane magnetization, while on the thicker side, the system preferred a phase with the antiferromagnetic out-of-plane order, as shown in Fig.~\ref{pic-pt}. 

As is expected from the theory of critical phenomena, the region of the film at the critical thickness showed strong dynamic magnetization fluctuations, which could be used to identify the spacial position of the boundary between the two phases. Due to the smoothness of the film thickness, the position of this boundary could be changed continuously by varying externally controlled parameters, such as temperature and the external magnetic field.
This property was used in  \cite{PhysRevX.8.031078}, in which the frequency-integrated magnetization noise power, proportional to the linear magnetic susceptibility, was measured in a fixed observation region. In a varying temperature, the critical point has passed through the observation spot and thus produced a peak of the integrated noise power as a function of the temperature. 

Unlike the susceptibility near a phase transition of a uniform system, this peak had a finite amplitude. In addition, Ref.~ \cite{PhysRevX.8.031078} revealed that the fluctuations at the critical values of the parameters were strongly non-Gaussian. They produce clearly resolvable bi-spectrum of fluctuations, which quickly disappears for the values of the control parameter away from the critical point. The enhanced fluctuations near a critical point have been later observed by means of spin noise spectroscopy in other magnetic systems, such as spin ice \cite{PhysRevX.11.011042}.

Here we would like to point out that the standard theory of critical phenomena cannot be directly applied in order to explain the fluctuations in the spatial critical region quantitatively. 
The reason is that when some parameter changes spatially, we do not deal with a uniform bulk sample. As the thickness in \cite{PhysRevB.90.184404} changes linearly along some axis $x$, the critical region in a $D$-dimensional sample degenerates to a $(D-1)$-dimensional surface. Hence, one cannot expect, e.g., the emergence of a divergent correlation length along the $x$-axis.  

On the other hand, some features of the theory of critical phenomena should remain valid near the spacial critical point. For example, if the thickness of the film changes substantially on much larger distances than the atomic lattice constant, then at the interface between the two phases, the magnetization fluctuations should also have much larger sizes than the lattice constant. Hence, the universality and scaling hypotheses from the theory of critical phenomena should still apply in some form to describe the experimental observations.

Thus, when a critical interface is set by imposing a weak spatial gradient of some parameter, the critical fluctuations along the gradient have essentially mesoscopic character: they are neither microscopic (decaying at the lattice constant scale) nor macroscopic (extending to the bulk of the sample). The question now is:  how to describe the physics near a critical point when some control parameter changes slowly and continuously throughout the critical point?

Optical magnetization noise spectroscopy has been proved to be an ideal tool to study such fluctuations experimentally because an optical probe beam covers a mesoscopic region of the sample and thus can capture the fluctuations specific to the critical region without a substantial contribution from the bulk of the sample. An additional question that we will explore is: what can we learn about the parameters of the system by observing the enhanced magnetization fluctuations near a spacial critical point? 
We will suggest a minimal model based on the Ginzburg-Landau effective free energy approach that contains the basic features of the problem. 


\section{Ginzburg-Landau Free Energy \label{sec:sec_II}}

Usually, phase transitions in quasi-1D systems can be understood with a free energy functional~\cite{Landau1937}
\begin{equation}\label{eq.1}
\mathcal{L}_{\mathrm{1D}} 
= 
\int_{-L}^L dx\, \left[\frac{D}{2}\left(\frac{d\phi}{dx}\right)^2 +\frac{r(x)}{2}\phi(x)^2+\frac{\vg}{4} \phi(x)^4\right].
\end{equation}
Here $\phi(x)$ represents a coarse-grained average order parameter. In the case of the ferromagnet-paramagnet phase transition, $\phi(x)$ is identified with the local magnetization $m_i = \braket{\hat{S^z_i}}$ in the original magnetic system (where $\mathbf{S}_i$'s denotes the spin degrees of freedom), $D$ is the diffusion coefficient, and $\vg>0$ is the coupling responsible for non-linearity associated with the mutual interaction between the individual spins. In the thermodynamic equilibrium, the ground state of the system is dictated by the minimum of $\mathcal{L}_{\mathrm{1D}}$. To make the phase transition well-defined, we assume a three-dimensional system along the transverse direction within a mesoscopic region of linear size much smaller than the quasi-1D length of the system.

We will assume that the phenomenological parameter $r$ is $x$-dependent. Without a diffusion term, the minimum of 
$\mathcal{L}_{\mathrm{1D}}$ is achieved at
\begin{subequations}
\begin{align}
\label{eq.2.1}
&   \phi    =   0,                      \qquad  \; \; \; \mathrm{for} \quad r \ge 0,   \\
\label{eq.2.2}
&   \phi    = \pm   \sqrt{\frac{r}{\vg}},     \quad \; \mathrm{for}		  \quad r < 0.
\end{align}
\end{subequations}
However, disregarding the diffusion, the calculation of the local susceptibility would give diverging predictions due to the discontinuity of the solution (\ref{eq.2.1})-(\ref{eq.2.2}) at the spacial point with $r=0$. Indeed, we will show that the properties of dynamic fluctuations near this point depend on the diffusion coefficient $D$ essentially.

In order to capture this physics, we include the diffusion term into consideration but use another simplification. Namely, near the critical point with $r=0$, we assume that 
the control parameter is only linearly varied along  $x$:
\begin{equation}\label{eq.3}
    r(x) = \theta x,
\end{equation}
where we set the critical point at $x=0$. 
The free energy functional in Eq.~(\ref{eq.1}) then modifies as
\begin{equation}\label{eq.4}
\scalebox{0.9}[1]{$\mathcal{L}_{\mathrm{1D}} 
= 
\int_{-L}^L dx  \left[\frac{D}{2}\left(\frac{d\phi}{dx}\right)^2  + \frac{\theta x}{2}\phi(x)^2 + \frac{\vg}{4} \phi(x)^4 - h\phi(x) \right],$}
\end{equation}
where we added an external ``magnetic field" $h$.

Far away from the critical point $x=0$, we have two phases: the phase-I for $ x>0$ with $\phi \rightarrow 0$ as $x\rightarrow +\infty$, and the phase-II for $x<0$ with $\phi(x) \rightarrow \pm \sqrt{\theta |x|/\vg}$ as $x \rightarrow -\infty$. In what follows, we will assume the symmetry to be broken in phase-II spontaneously so that
\begin{subequations}
\begin{align}
\label{eq.5.1}
&   \phi(x) \rightarrow \sqrt{\theta |x|/\vg},     \quad 			   		x	\rightarrow -\infty, \\
\label{eq.5.2}
&   \phi(x) \rightarrow 0,                      \qquad \quad \quad 		x 	\rightarrow +\infty.    
\end{align}
\end{subequations}
To find the equilibrium $\phi(x)$, we should minimize the new functional in Eq.~(\ref{eq.4}). This gives us an equation
\begin{equation}\label{eq.6}
-D\frac{d^2 \phi}{dx^2} + \theta x \phi(x) + \vg \phi(x)^3 = h.
\end{equation}
Note that by dividing this equation by $D$, we find that only  the combinations, $\theta/D$, $g/D$, and $h/D$ matter. Then, by rescaling space, $x\rightarrow \lambda x$, we get the same equation but with $\theta \rightarrow \theta \lambda^3$ and $\vg \rightarrow \vg \lambda^2$. Choosing 
\begin{equation}\label{eq.7}
\lambda     =   (D/\theta)^{1/3}, 
\end{equation}
we get rid of the coefficient near $-x\phi(x)$. The coefficient $\vg$ then becomes $\vg \rightarrow \vg /(D \theta^{2})^{1/3}$. Finally, by rescaling $\phi \rightarrow \mu \phi$, we can set the coefficient near $\phi^3$ to be equal to 2. This is achieved for 
\begin{equation}\label{eq.8}
\mu =\sqrt{\frac{2D}{\vg \lambda^2}}=\frac{\sqrt{2}D^{1/6}\theta^{1/3}}{\vg^{1/2}}.
\end{equation}
In terms of these scaled variables and parameters, the differential Eq.~(\ref{eq.6}) is modified with a rescaled average local magnetization, $u(x,\alpha)$, satisfying 
\begin{equation}\label{eq.9}
\frac{d^2 u(x,\alpha)}{dx^2}  =   2 u(x,\alpha)^3 + x u(x,\alpha)  - \alpha,
\end{equation}
where
\begin{equation}\label{eq.10}
\alpha  =   h   \frac{\lambda^2}{D \mu } = \frac{h \sqrt{\vg} }{\theta \sqrt{2D}}.
\end{equation}
Then, the solution of the original Eq.~(\ref{eq.6}) is given by
\begin{equation}\label{eq.11}
    \phi(x|h) = \mu u( x/\lambda,\alpha(h)).
\end{equation}
Note that this reduction predicts the characteristic scales for the  magnetic field
\begin{equation}\label{eq.12}
h_0
=
\frac{\theta \sqrt{2D}}{\sqrt{g}}, \quad \alpha  =   h/h_0,
\end{equation}
and also the length $\lambda$ in Eq.~(\ref{eq.7}), as well as the magnetization $\mu$ in Eq.~(\ref{eq.8}). All these scales depend on the diffusion coefficient $D$ and the slope $\theta$. Hence, they must be related to the observable physics of the magnetic fluctuations within the length $\lambda$ near the critical point.

\section{Solution of the Painleve equation \label{sec:sec_III}}

Equation~\eqref{eq.9} is known as the Painlev\'e-II ($\mathrm{P}_{\mathrm{II}}$) non-linear differential equation. The fact that this equation can be obtained by minimizing a Ginzburg-Landau functional is well known~\cite{Conte1993,Profilo1991,Liu2009}. However, here our interest is more specific.

Namely, spin noise spectroscopy generally measures the noise power spectra, whose integrals over frequency return linear and nonlinear susceptibilities of a mesoscopic probed region. 
The solution of Painlev\'e-II is generally considered as independent special function because many of its properties are known analytically. In particular,  very recently, it was shown that the integrals of the Painlev\'e-II solutions could be written analytically in terms of standard special functions \cite{DAI20202430}. 
A real solution of Eq.~(\ref{eq.9}) that would be consistent with the conditions Eqs.~(\ref{eq.5.1})-(\ref{eq.5.2}) is unique and known as the Hastings-McLeod solution~\cite{Hastings1980}.
Its asymptotic behavior as $x\rightarrow \pm \infty$ and $\alpha \in (-1/2,1/2)$ is also well known \cite{Dai_2017}:
\begin{widetext}
\begin{subequations}
\begin{align}
\label{eq.13.1}
    u(x;\alpha)_{x\rightarrow +\infty} &    =   B(\alpha;x)+\cos(\pi \alpha)({\rm Ai}(x) + \mathcal{O}(x^{-3/4})), \\
\label{eq.13.2}
    u(x;\alpha)_{x\rightarrow -\infty} &    =   \sqrt{-x/2} \left(1+\frac{\alpha}{\sqrt{2}|x|^{3/2}}-\frac{1+6\alpha^2}{8|x|^3} + \mathcal{O}(1/|x|^{9/2}) \right)    
                                                +   [\mathrm{exponentially} \,\, \mathrm{small} \,\, \mathrm{terms}], \\
\label{eq.13.3}
    B(\alpha;x)                        &    =   \frac{\alpha}{x} \sum_{n=0}^{\infty} \frac{a_n}{x^{3n}}, \qquad
    a_0 = 1, \;
    a_{n+1} = (3n+1)(3n+2)a_n - 2\alpha^2 \sum_{k,l,m=0}^n a_k a_l a_m.
\end{align}
\end{subequations}
\end{widetext}
At this stage, we note that the asymptotic behavior depends on two types of terms, i.e., the terms that decay as a power law and the exponentially suppressed terms. 

Examination of the leading order power-law decay terms shows that it is not related to the intrinsic to $x=0$ physics. Namely, the leading asymptotic as $x\rightarrow -\infty$ in Eq.~(\ref{eq.13.2}) corresponds to the $\phi \rightarrow \sqrt{|x|}$ behavior, which is expected from Eq.~(\ref{eq.5.1}). For $x\rightarrow+\infty$, the term $B(\alpha;x)$ becomes zero at $\alpha=0$, i.e., it describes a diamagnetic response in the nonmagnetic phase. Its appearance is expected because, for large $x$ and small $\phi(x)$, we can disregard the diffusion and nonlinear contributions so that the energy density becomes $ \mathcal{E}(x) \sim \theta x \phi(x)^2 - h \phi(x)$, which is minimized at $\phi \sim h/(\theta x)$, and which reproduces the leading asymptotic in Eq.~(\ref{eq.13.3}) if we use it in the solution in Eq.~(\ref{eq.11}).

The other terms are decaying exponentially. They play an important role for $\alpha=0$, \textit{i.e.} in the absence of the external magnetic field [see Eq.~(\ref{eq.10})]. The limit $x\rightarrow +\infty$ is then dominated by the Airy function in Eq.~(\ref{eq.5.1}), which leads to the asymptotic
\begin{equation}\label{eq.14}
	u(x;\alpha=0)_{x\rightarrow +\infty} 	\sim	\frac{x^{-1/4}}{2\sqrt{3\pi}}e^{-\frac{2}{3}|x|^{3/2}}.
\end{equation}
Returning to the non-rescaled variables, we find 
\begin{equation}\label{eq.15}
	\phi(x;h=0)_{x\rightarrow +\infty} 		\sim	\frac{\mu (x/l)^{-1/4}}{2\sqrt{3\pi}}e^{-\frac{2}{3}(|x|/\lambda)^{3/2}}.
\end{equation}
Thus, we find that the magnetization does not drop to zero instantly in phase-I but rather persists up to the distance $l=(D/\theta)^{1/3}$ from the critical point with a characteristic amplitude $\mu$, and only then decays faster than exponentially with growing $x$. This means that the behavior near the critical point is dominated by the proximity effect, in which extension and amplitude are influenced by both the diffusion $D$ and the ramp of the transition $\theta$. Hence, already at the level of the asymptotic analysis, we find that the behavior of the magnetization contains the terms that are strongly localized near the critical point and, therefore, can be attributed only to the physics in its vicinity. 

\section{Linear and nonlinear susceptibilities of a mesoscopic critical region \label{sec:sec_IV}}

\subsection{Linear susceptibility far from critical point \label{sec:sec_IV.I}}

Spin noise spectroscopy is the method that allows studies of the local spin susceptibilities, which are averaged over a mesoscopic region in space \cite{Sinitsyn_2016}. 
This technique can generally resolve the full frequency dependence of magnetization time-correlators. However, the ground state configuration of the magnetic system, obtained by minimizing $\mathcal{L}_{\mathrm{1D}}$ as in Eq.~(\ref{eq.6}), contains information only about the static  characteristics, such as the local linear and the nonlinear susceptibilities:
\begin{equation}\label{eq.16}
\chi^{(1)}_x
=
\left(\frac{\partial \phi(x)}{\partial h}\right)_{h=0},
\quad
\chi^{(2)}_x
=
\frac{1}{2}\left(\frac{\partial^2 \phi(x)}{\partial h^2}\right)_{h=0}, \cdots   
\end{equation}
and which are obtained experimentally by calculating the spectral volumes of the measured correlators. 

Due to the finite size of the optical beam width ($L>1\mu$m), the measured magnetization is averaged over the length $L>\lambda$. If we are far from the point $x=0$, the order parameter changes slowly with $x$. So, asymptotically we find the local susceptibility [from Eq.~(\ref{eq.13.1}), and Eq.~(\ref{eq.13.2})]: 
\begin{subequations}
\begin{align}
\label{eq.17.1}
	\chi^{(1)}_x 	&	\propto \frac{1}{\theta x}, \quad x \rightarrow +\infty, \\
\label{eq.17.2}	
	\chi^{(1)}_x 	&	\propto \frac{1}{2\theta x}, \quad x \rightarrow -\infty.
\end{align}
\end{subequations}
which is  a typical $\lambda$-like power-law behavior, near a critical point. However, this behavior is not related to a specific physics near the critical point. Moreover, it shows that the linear local susceptibility is likely not a good characteristic to explore these physics because the integrated over the macroscopic region susceptibility diverges with the growing width of the probe beam. 
Indeed, let the probe beam cut the susceptibility from the region $x\in (-c,s)$, where $c$ and $s$ are much larger than $\lambda$. Then, the integrated susceptibility
\begin{equation}\label{eq.18}
	\chi^{(1)}_{\mathrm{total}}	=	\int_{-c}^s \chi^{(1)}_x \, dx.
\end{equation}
behaves as
\begin{equation}\label{eq.19}
	\chi^{(1)}_{\mathrm{total}} 	\sim \chi_0 + \frac{1}{\theta} \left( \frac{\ln c}{2} + \ln s \right),
\end{equation}
where $\chi_0$ is some constant that depends on the intermediate profile of $\chi^{(1)}_x$. This expression is dominated by the logarithmic tails with the probe-dependent cutoff. We will show that the constant part of such an integrated susceptibility is not very valuable either.

\subsection{Excess magnetization and linear susceptibility \label{sec:sec_IV.II}}

Let us define the net magnetization in the observation region
\begin{equation}\label{eq.22}
    M={\rm lim}_{c,s \rightarrow \infty} \int_{-c}^s \phi(x)\, dx. 
\end{equation}
Using Eq.~(\ref{eq.11}) we can rewrite
\begin{equation}\label{eq.23}
	M	=	\mu \lambda\int_{-c'}^{s'}u(x, \alpha(h))\, dx,
\end{equation}
where $c'=c/\lambda$, $s'=s/\lambda$. The integrated susceptibilities are then defined as
\begin{equation}\label{eq.24}
	\chi^{(n)}_{\mathrm{total}}	 =  \frac{1}{n!}\left(\frac{\partial^n M}{\partial h^n}\right)_{h=0} =\frac{1}{n!}\frac{\mu \lambda}{h_0^n} \left(\frac{\partial^n {\cal M}}{\partial \alpha^n}\right)_{\alpha=0}, 
\end{equation}
where 
\begin{equation}\label{eq.25}
	\mathcal{M}	=	{\rm lim}_{c',s' \rightarrow \infty} \int_{-c'}^{s'} u(x,\alpha)\, dx.
\end{equation}
According to \cite{DAI20202430}, this integral is known:
\begin{widetext}
\begin{equation}\label{eq.26}
    \mathcal{M} = \frac{\sqrt{2}c'^{3/2}}{3} + 
    \frac{\alpha}{2} \ln c' + \alpha \ln s' +\frac{1}{2} \ln (2\pi) - \ln \Gamma(\alpha+1/2) -\frac{\alpha \ln 2}{2}+O(1/s'^{3/2},1/c'^{3/2}).
\end{equation}
\end{widetext}
Setting $\alpha =0$ in Eq.~(\ref{eq.26}), and using that $\mu\lambda = \sqrt{2D/\vg}$, we find the net magnetization
\begin{equation}\label{eq.27}
	M	=	\sqrt{\frac{2D}{\vg}} \left(\frac{\sqrt{2} c'^{3/2}}{3} +\frac{ \ln 2}{2} \right). 
\end{equation}
The term $\sim c'^{3/2}$ is expected from the diffusionless limit. Indeed, using that $c'=c/\lambda$, and $\lambda^{3/2} \sim \sqrt{D}$, we find that this term does not depend on the diffusion coefficient. The contribution
\begin{equation}\label{eq.28}
	\delta M 	=	\sqrt{\frac{D}{2\vg}} \ln 2,
\end{equation}
is, however, a correction due to the proximity effects near the critical point. Interestingly, this ``excess magnetization" correction does not depend on the slope $\theta$, which is our first prediction for the property of the critical region. 

Potentially, $\delta M$ can be measured optically by scanning the dependence of the average Kerr rotation on the area of the probe beam and fitting the result to a power law with an offset. However, on the background of the dominating trivial $\sim c'^{3/2}$ contribution, this would be hard. Hence, let us look now at the linear susceptibility. Using that $\mu \lambda /h_0=1/\theta$, we find
\begin{equation}\label{eq.29}
\chi^{(1)}_{\mathrm{total}} 
=
\frac{1}{\theta} \bigg[ \frac{1}{2} \ln c' + \ln s' +\left( \gamma_e +\frac{3 \ln 2}{2} \right) \bigg]  
\end{equation}
where $\gamma_e\approx 0.577$ is the Euler–Mascheroni constant. This expression shows that the integrated linear susceptibility of the critical region is quite uninformative. In addition to the logarithmic cutoff-dependent terms, it has a subdominant contribution that depends only on the gradient of the control parameter, $\theta$. 

Still, this susceptibility is a valuable pointer to the critical point, as it reaches the maximum when this point is placed inside the integration interval $(-c,s)$.
However, as a small warning to experiments, we note that within our model, the factor $1/2$ in (\ref{eq.29}) leads to a mismatch of the critical point from the center of this interval. For example, if the width of the interval is $L$, then $s=L-c$, and the maximum of the linear susceptibility is found at $c=L/3$ rather than $L/2$.


\subsection{Non-linear susceptibilities \label{sec:sec_IV.IV}}

The behavior that is intrinsic only to the critical region is found in all higher-order susceptibilities. Thus
\begin{subequations}
\begin{align}
\label{eq.30.1}
\chi^{(2)}_{\mathrm{total}}
& =
-\frac{\pi^2 \mu \lambda}{4h_0} =-\frac{\pi^2 \sqrt{\vg}}{4\theta^2 \sqrt{2D}}, \\
\label{eq.30.2}
\chi^{(3)}_{\mathrm{total}}
& =
\frac{7 \zeta(3) \mu \lambda}{3h_0^3}=\frac {7\zeta(3)\vg}{6\theta^3 D },
\end{align}
\end{subequations}
where $\zeta(3) \approx 1.20$ is a special value of the Riemann zeta-function. First, we note the absence of the cutoff-dependent contributions to $\chi^{(2)}_{\mathrm{total}}$ and $\chi^{(3)}_{\mathrm{total}}$. This means that they are dominated by the physics near the critical point. Indeed, far away from the critical point, the magnetic fluctuations are essentially Gaussian because they are dominated by numerous uncorrelated microscopic events. Near the critical point, the fluctuations are not only enhanced, but they also become highly non-Gaussian. Therefore,  the nonlinear susceptibilities acquire finite contributions precisely near the critical region, in agreement with the experimental observation in \cite{PhysRevX.8.031078}. 

 In addition, we note the singular dependence of the nonlinear susceptibilities on the diffusion coefficient $D$. Due to the critical slowing-down, the fluctuations at the critical point have a formally infinite lifetime, which would lead to a diverging bi-spectra near the frequency $\omega_{1,2}=0$ point. However, the diffusion introduces a new characteristic lifetime that smears this divergence and makes the integrated susceptibility finite.
Such behavior is consistent with the appearance of $D$ in the denominators of (\ref{eq.30.1}) and (\ref{eq.30.2}).

\section{Discussion \label{sec:sec_V}}

Our theoretical findings confirm the experimental observation of the growing role of the non-Gaussian fluctuations within the mesoscopic critical region. We proposed the minimal model that demonstrates this effect and can characterize it quantitatively. 

We showed that the nonlinear susceptibilities
 contain information about the intrinsic parameters of the interacting spins in the critical region.
We also predict a quite universal behavior of the integrated linear susceptibility and the excess magnetization of the mesoscopic critical region. The theory also reveals the power-law dependence of the measurable susceptibilities on the gradient of the control parameter $\theta$. All this points to the possible universality of such fluctuations across many different systems.


Based on our results, we conjecture that, due to the presence of long-range correlations, there should generally be effective field theories describing such critical regions, possibly with a few most important universality classes, as in the theory of standard critical phenomena. As in our case, such effective theories might be integrable. Due to the broken chiral symmetry induced by a gradient of the parameter, they also may have exotic topologically protected excitations. Finally, there can be practical applications for experimental studies of such regions. For example, they are movable within the samples by changing externally controlled parameters, such as the temperature and the external fields. This may be used for adiabatic transport of exotic quasiparticles or inducing large magnetic fluctuations on demand locally.

\section{Acknowledgment \label{sec:sec_VI}}

This work was supported by the U.S. Department of Energy (DOE), Office of Science, and Office of Advanced Scientific Computing Research, through the Quantum Internet to Accelerate Scientific Discovery Program. 

\bibliography{References}

\begin{thebibliography}{15}%
\makeatletter
\providecommand \@ifxundefined [1]{%
 \@ifx{#1\undefined}
}%
\providecommand \@ifnum [1]{%
 \ifnum #1\expandafter \@firstoftwo
 \else \expandafter \@secondoftwo
 \fi
}%
\providecommand \@ifx [1]{%
 \ifx #1\expandafter \@firstoftwo
 \else \expandafter \@secondoftwo
 \fi
}%
\providecommand \natexlab [1]{#1}%
\providecommand \enquote  [1]{``#1''}%
\providecommand \bibnamefont  [1]{#1}%
\providecommand \bibfnamefont [1]{#1}%
\providecommand \citenamefont [1]{#1}%
\providecommand \href@noop [0]{\@secondoftwo}%
\providecommand \href [0]{\begingroup \@sanitize@url \@href}%
\providecommand \@href[1]{\@@startlink{#1}\@@href}%
\providecommand \@@href[1]{\endgroup#1\@@endlink}%
\providecommand \@sanitize@url [0]{\catcode `\\12\catcode `\$12\catcode
  `\&12\catcode `\#12\catcode `\^12\catcode `\_12\catcode `\%12\relax}%
\providecommand \@@startlink[1]{}%
\providecommand \@@endlink[0]{}%
\providecommand \url  [0]{\begingroup\@sanitize@url \@url }%
\providecommand \@url [1]{\endgroup\@href {#1}{\urlprefix }}%
\providecommand \urlprefix  [0]{URL }%
\providecommand \Eprint [0]{\href }%
\providecommand \doibase [0]{http://dx.doi.org/}%
\providecommand \selectlanguage [0]{\@gobble}%
\providecommand \bibinfo  [0]{\@secondoftwo}%
\providecommand \bibfield  [0]{\@secondoftwo}%
\providecommand \translation [1]{[#1]}%
\providecommand \BibitemOpen [0]{}%
\providecommand \bibitemStop [0]{}%
\providecommand \bibitemNoStop [0]{.\EOS\space}%
\providecommand \EOS [0]{\spacefactor3000\relax}%
\providecommand \BibitemShut  [1]{\csname bibitem#1\endcsname}%
\let\auto@bib@innerbib\@empty
\bibitem [{\citenamefont {Balk}\ \emph {et~al.}(2014)\citenamefont {Balk},
  \citenamefont {Stiles},\ and\ \citenamefont {Unguris}}]{PhysRevB.90.184404}%
  \BibitemOpen
  \bibfield  {author} {\bibinfo {author} {\bibfnamefont {A.~L.}\ \bibnamefont
  {Balk}}, \bibinfo {author} {\bibfnamefont {M.~D.}\ \bibnamefont {Stiles}}, \
  and\ \bibinfo {author} {\bibfnamefont {J.}~\bibnamefont {Unguris}},\ }\href
  {\doibase 10.1103/PhysRevB.90.184404} {\bibfield  {journal} {\bibinfo
  {journal} {Phys. Rev. B}\ }\textbf {\bibinfo {volume} {90}},\ \bibinfo
  {pages} {184404} (\bibinfo {year} {2014})}\BibitemShut {NoStop}%
\bibitem [{\citenamefont {Im}\ \emph {et~al.}(2009)\citenamefont {Im},
  \citenamefont {Fischer}, \citenamefont {Kim},\ and\ \citenamefont
  {Shin}}]{Young2009}%
  \BibitemOpen
  \bibfield  {author} {\bibinfo {author} {\bibfnamefont {M.-Y.}\ \bibnamefont
  {Im}}, \bibinfo {author} {\bibfnamefont {P.}~\bibnamefont {Fischer}},
  \bibinfo {author} {\bibfnamefont {D.-H.}\ \bibnamefont {Kim}}, \ and\
  \bibinfo {author} {\bibfnamefont {S.-C.}\ \bibnamefont {Shin}},\ }\href
  {\doibase 10.1063/1.3256188} {\bibfield  {journal} {\bibinfo  {journal}
  {Appl. Phys. Lett.}\ }\textbf {\bibinfo {volume} {95}} (\bibinfo {year}
  {2009}),\ 10.1063/1.3256188},\ \bibinfo {note} {182504}\BibitemShut {NoStop}%
\bibitem [{\citenamefont {Christian}\ \emph {et~al.}(2006)\citenamefont
  {Christian}, \citenamefont {Novoselov},\ and\ \citenamefont
  {Geim}}]{PhysRevB.74.064403}%
  \BibitemOpen
  \bibfield  {author} {\bibinfo {author} {\bibfnamefont {D.~A.}\ \bibnamefont
  {Christian}}, \bibinfo {author} {\bibfnamefont {K.~S.}\ \bibnamefont
  {Novoselov}}, \ and\ \bibinfo {author} {\bibfnamefont {A.~K.}\ \bibnamefont
  {Geim}},\ }\href {\doibase 10.1103/PhysRevB.74.064403} {\bibfield  {journal}
  {\bibinfo  {journal} {Phys. Rev. B}\ }\textbf {\bibinfo {volume} {74}},\
  \bibinfo {pages} {064403} (\bibinfo {year} {2006})}\BibitemShut {NoStop}%
\bibitem [{\citenamefont {Liebmann}\ \emph {et~al.}(2005)\citenamefont
  {Liebmann}, \citenamefont {Schwarz}, \citenamefont {Kaiser}, \citenamefont
  {Wiesendanger}, \citenamefont {Kim},\ and\ \citenamefont
  {Noh}}]{PhysRevB.71.104431}%
  \BibitemOpen
  \bibfield  {author} {\bibinfo {author} {\bibfnamefont {M.}~\bibnamefont
  {Liebmann}}, \bibinfo {author} {\bibfnamefont {A.}~\bibnamefont {Schwarz}},
  \bibinfo {author} {\bibfnamefont {U.}~\bibnamefont {Kaiser}}, \bibinfo
  {author} {\bibfnamefont {R.}~\bibnamefont {Wiesendanger}}, \bibinfo {author}
  {\bibfnamefont {D.-W.}\ \bibnamefont {Kim}}, \ and\ \bibinfo {author}
  {\bibfnamefont {T.-W.}\ \bibnamefont {Noh}},\ }\href {\doibase
  10.1103/PhysRevB.71.104431} {\bibfield  {journal} {\bibinfo  {journal} {Phys.
  Rev. B}\ }\textbf {\bibinfo {volume} {71}},\ \bibinfo {pages} {104431}
  (\bibinfo {year} {2005})}\BibitemShut {NoStop}%
\bibitem [{\citenamefont {Shin}\ \emph {et~al.}(2007)\citenamefont {Shin},
  \citenamefont {Ryu}, \citenamefont {Kim}, \citenamefont {Choe},\ and\
  \citenamefont {Akinaga}}]{Shin2007}%
  \BibitemOpen
  \bibfield  {author} {\bibinfo {author} {\bibfnamefont {S.-C.}\ \bibnamefont
  {Shin}}, \bibinfo {author} {\bibfnamefont {K.-S.}\ \bibnamefont {Ryu}},
  \bibinfo {author} {\bibfnamefont {D.-H.}\ \bibnamefont {Kim}}, \bibinfo
  {author} {\bibfnamefont {S.-B.}\ \bibnamefont {Choe}}, \ and\ \bibinfo
  {author} {\bibfnamefont {H.}~\bibnamefont {Akinaga}},\ }\href {\doibase
  https://doi.org/10.1016/j.jmmm.2006.11.041} {\bibfield  {journal} {\bibinfo
  {journal} {Journal of Magnetism and Magnetic Materials}\ }\textbf {\bibinfo
  {volume} {310}},\ \bibinfo {pages} {2599} (\bibinfo {year} {2007})},\
  \bibinfo {note} {proceedings of the 17th International Conference on
  Magnetism}\BibitemShut {NoStop}%
\bibitem [{\citenamefont {Balk}\ \emph {et~al.}(2018)\citenamefont {Balk},
  \citenamefont {Li}, \citenamefont {Gilbert}, \citenamefont {Unguris},
  \citenamefont {Sinitsyn},\ and\ \citenamefont {Crooker}}]{PhysRevX.8.031078}%
  \BibitemOpen
  \bibfield  {author} {\bibinfo {author} {\bibfnamefont {A.~L.}\ \bibnamefont
  {Balk}}, \bibinfo {author} {\bibfnamefont {F.}~\bibnamefont {Li}}, \bibinfo
  {author} {\bibfnamefont {I.}~\bibnamefont {Gilbert}}, \bibinfo {author}
  {\bibfnamefont {J.}~\bibnamefont {Unguris}}, \bibinfo {author} {\bibfnamefont
  {N.~A.}\ \bibnamefont {Sinitsyn}}, \ and\ \bibinfo {author} {\bibfnamefont
  {S.~A.}\ \bibnamefont {Crooker}},\ }\href {\doibase
  10.1103/PhysRevX.8.031078} {\bibfield  {journal} {\bibinfo  {journal} {Phys.
  Rev. X}\ }\textbf {\bibinfo {volume} {8}},\ \bibinfo {pages} {031078}
  (\bibinfo {year} {2018})}\BibitemShut {NoStop}%
\bibitem [{\citenamefont {Goryca}\ \emph {et~al.}(2021)\citenamefont {Goryca},
  \citenamefont {Zhang}, \citenamefont {Li}, \citenamefont {Balk},
  \citenamefont {Watts}, \citenamefont {Leighton}, \citenamefont {Nisoli},
  \citenamefont {Schiffer},\ and\ \citenamefont
  {Crooker}}]{PhysRevX.11.011042}%
  \BibitemOpen
  \bibfield  {author} {\bibinfo {author} {\bibfnamefont {M.}~\bibnamefont
  {Goryca}}, \bibinfo {author} {\bibfnamefont {X.}~\bibnamefont {Zhang}},
  \bibinfo {author} {\bibfnamefont {J.}~\bibnamefont {Li}}, \bibinfo {author}
  {\bibfnamefont {A.~L.}\ \bibnamefont {Balk}}, \bibinfo {author}
  {\bibfnamefont {J.~D.}\ \bibnamefont {Watts}}, \bibinfo {author}
  {\bibfnamefont {C.}~\bibnamefont {Leighton}}, \bibinfo {author}
  {\bibfnamefont {C.}~\bibnamefont {Nisoli}}, \bibinfo {author} {\bibfnamefont
  {P.}~\bibnamefont {Schiffer}}, \ and\ \bibinfo {author} {\bibfnamefont
  {S.~A.}\ \bibnamefont {Crooker}},\ }\href {\doibase
  10.1103/PhysRevX.11.011042} {\bibfield  {journal} {\bibinfo  {journal} {Phys.
  Rev. X}\ }\textbf {\bibinfo {volume} {11}},\ \bibinfo {pages} {011042}
  (\bibinfo {year} {2021})}\BibitemShut {NoStop}%
\bibitem [{\citenamefont {Landau}(1937)}]{Landau1937}%
  \BibitemOpen
  \bibfield  {author} {\bibinfo {author} {\bibfnamefont {L.~D.}\ \bibnamefont
  {Landau}},\ }\href
  {http://archive.ujp.bitp.kiev.ua/files/journals/53/si/53SI08p.pdf} {\bibfield
   {journal} {\bibinfo  {journal} {Zh. Eksp. Teor. Fiz.}\ }\textbf {\bibinfo
  {volume} {7}},\ \bibinfo {pages} {19} (\bibinfo {year} {1937})}\BibitemShut
  {NoStop}%
\bibitem [{\citenamefont {Conte}\ and\ \citenamefont
  {Musette}(1993)}]{Conte1993}%
  \BibitemOpen
  \bibfield  {author} {\bibinfo {author} {\bibfnamefont {R.}~\bibnamefont
  {Conte}}\ and\ \bibinfo {author} {\bibfnamefont {M.}~\bibnamefont
  {Musette}},\ }\href {\doibase https://doi.org/10.1016/0167-2789(93)90177-3}
  {\bibfield  {journal} {\bibinfo  {journal} {Phys. D: Nonlinear Phenom.}\
  }\textbf {\bibinfo {volume} {69}},\ \bibinfo {pages} {1} (\bibinfo {year}
  {1993})}\BibitemShut {NoStop}%
\bibitem [{\citenamefont {Profilo}\ and\ \citenamefont
  {Soliani}(1991)}]{Profilo1991}%
  \BibitemOpen
  \bibfield  {author} {\bibinfo {author} {\bibfnamefont {G.}~\bibnamefont
  {Profilo}}\ and\ \bibinfo {author} {\bibfnamefont {G.}~\bibnamefont
  {Soliani}},\ }\href {\doibase 10.1007/BF02759774} {\bibfield  {journal}
  {\bibinfo  {journal} {II Nuovo Cim. B}\ }\textbf {\bibinfo {volume} {106}},\
  \bibinfo {pages} {307} (\bibinfo {year} {1991})}\BibitemShut {NoStop}%
\bibitem [{\citenamefont {Liu}\ \emph {et~al.}(2009)\citenamefont {Liu},
  \citenamefont {Li},\ and\ \citenamefont {Tian}}]{Liu2009}%
  \BibitemOpen
  \bibfield  {author} {\bibinfo {author} {\bibfnamefont {J.-G.}\ \bibnamefont
  {Liu}}, \bibinfo {author} {\bibfnamefont {Y.-Z.}\ \bibnamefont {Li}}, \ and\
  \bibinfo {author} {\bibfnamefont {B.}~\bibnamefont {Tian}},\ }\href {\doibase
  https://doi.org/10.1016/j.cnsns.2008.01.011} {\bibfield  {journal} {\bibinfo
  {journal} {Commun. Nonlinear Sci. Numer. Simul.}\ }\textbf {\bibinfo {volume}
  {14}},\ \bibinfo {pages} {1214} (\bibinfo {year} {2009})}\BibitemShut
  {NoStop}%
\bibitem [{\citenamefont {Dai}\ \emph {et~al.}(2020)\citenamefont {Dai},
  \citenamefont {Xu},\ and\ \citenamefont {Zhang}}]{DAI20202430}%
  \BibitemOpen
  \bibfield  {author} {\bibinfo {author} {\bibfnamefont {D.}~\bibnamefont
  {Dai}}, \bibinfo {author} {\bibfnamefont {S.-X.}\ \bibnamefont {Xu}}, \ and\
  \bibinfo {author} {\bibfnamefont {L.}~\bibnamefont {Zhang}},\ }\href
  {\doibase https://doi.org/10.1016/j.jde.2020.02.003} {\bibfield  {journal}
  {\bibinfo  {journal} {J. Differ. Equ.}\ }\textbf {\bibinfo {volume} {269}},\
  \bibinfo {pages} {2430} (\bibinfo {year} {2020})}\BibitemShut {NoStop}%
\bibitem [{\citenamefont {Hastings}\ and\ \citenamefont
  {McLeod}(1980)}]{Hastings1980}%
  \BibitemOpen
  \bibfield  {author} {\bibinfo {author} {\bibfnamefont {S.~P.}\ \bibnamefont
  {Hastings}}\ and\ \bibinfo {author} {\bibfnamefont {J.~B.}\ \bibnamefont
  {McLeod}},\ }\href {\doibase 10.1007/BF00283254} {\bibfield  {journal}
  {\bibinfo  {journal} {Arch. Ration. Mech. Anal.}\ }\textbf {\bibinfo {volume}
  {73}},\ \bibinfo {pages} {31} (\bibinfo {year} {1980})}\BibitemShut {NoStop}%
\bibitem [{\citenamefont {Dai}\ and\ \citenamefont {Hu}(2017)}]{Dai_2017}%
  \BibitemOpen
  \bibfield  {author} {\bibinfo {author} {\bibfnamefont {D.}~\bibnamefont
  {Dai}}\ and\ \bibinfo {author} {\bibfnamefont {W.}~\bibnamefont {Hu}},\
  }\href {\doibase 10.1088/1361-6544/aa72c1} {\bibfield  {journal} {\bibinfo
  {journal} {Nonlinearity}\ }\textbf {\bibinfo {volume} {30}},\ \bibinfo
  {pages} {2982} (\bibinfo {year} {2017})}\BibitemShut {NoStop}%
\bibitem [{\citenamefont {Sinitsyn}\ and\ \citenamefont
  {Pershin}(2016)}]{Sinitsyn_2016}%
  \BibitemOpen
  \bibfield  {author} {\bibinfo {author} {\bibfnamefont {N.~A.}\ \bibnamefont
  {Sinitsyn}}\ and\ \bibinfo {author} {\bibfnamefont {Y.~V.}\ \bibnamefont
  {Pershin}},\ }\href {\doibase 10.1088/0034-4885/79/10/106501} {\bibfield
  {journal} {\bibinfo  {journal} {Rep. Prog. Phys.}\ }\textbf {\bibinfo
  {volume} {79}},\ \bibinfo {pages} {106501} (\bibinfo {year}
  {2016})}\BibitemShut {NoStop}%
\end{thebibliography}%
\bibliographystyle{apsrev4-1}

\end{document}